\documentclass[review,authoryear]{elsarticle}
\journal{Transportation Research Part B: Methodological}
\pdfoutput=1
\usepackage{amsmath,amssymb,bm}
\usepackage{graphics}
\usepackage[pdftex,colorlinks,bookmarks,pdfpagemode=UseOutlines,linkcolor=black,urlcolor=black,citecolor=black,pageanchor=false]{hyperref}
\usepackage[utf8]{inputenc}
\usepackage[english]{babel}

\usepackage{url}
\usepackage{epstopdf}

\usepackage{amssymb,amsmath}

\usepackage{microtype}

\title{Derivation, calibration and verification of macroscopic model for urban traffic flow. Part~2}

\author[mipt,icad]{A.E.~Alekseenko\corref{cor1}}
\ead{aland@phystech.edu}
\cortext[cor1]{Corresponding author, \url{aland@phystech.edu}}
\author[mipt,icad,inno]{Ya.A.~Kholodov}
\author[mipt,icad]{A.S.~Kholodov}
\author[mipt]{A.I.~Goreva}
\author[ucb]{A.A.~Kurzhanskiy}
\author[mipt,ccas]{Yu.V.~Chehovich}
\author[mipt]{V.M.~Starozhilets}

\address[mipt]{Moscow Institute of~Physics and~Technology, Dolgoprudny, Moscow Region, 141707, Russia}
\address[icad]{Institute for Computer Aided Design of~RAS, Moscow, 123056, Russia}
\address[inno]{Innopolis University, Innopolis, Tatarstan, 420500, Russia}
\address[ccas]{Dorodnicyn Computing Centre of~RAS, Moscow, 119333, Russia}
\address[ucb]{University of~California, Berkeley, CA 94720-1770, United States}

\begin{document}

\begin{abstract}
In this paper, we propose a unified procedure for calibration of 
macroscopic second-order multilane traffic models.
The focus is on calibrating the fundamental diagram using the combination
stationary detector data and GPS traces.
GPS traces are used in estimation of the deceleration wave speed.
Thus calibrated model adequately represents the three phases of traffic:
free flow, synchronized flow and the wide moving jam.
The proposed approach was validated in simulation using stationary
detection data and GPS traces from the Moscow Ring Road.
Simulation showed that the proposed second-order model is more accurate
than the first-order LWR model.

The work was supported by grant RSCF~14-11-00877.
\end{abstract}

\begin{keyword}
traffic flow \sep macroscopic hydrodynamical models \sep phenomenological models \sep computer vision \sep fundamental diagram
\end{keyword}

\maketitle

\section{Introduction}
In three-phase traffic theory~\citep{Kerner2009}, the distinction between
free and congested flow is the same as in classical
Lighthill--Whitham--Richards (LWR)~\citep{Lighthill1955,Richards1956,Whitham1974} and General Motors (GM) \citep{Gazis1974} models. 
The fundamental feature of Kerner's model is the division of congested state
into two phases based on empirical properties of traffic flow,
which has been consistently observed on numerous roadways 
in different countries.
In Part~1 of this paper we introduced the second-order hyperbolic PDE model
that captures the three-phase nature of traffic flow.

Today, the combination of GPS traces from vehicles and data from
stationary detectors allow us to estimate more parameters and, thus,
develop more elaborate models.
The quality of model parameter estimation, however, heavily depends
on completeness of input data.
In this part of the paper we show how the combination of measurements from
both stationary and moving sensors is used to calibrate the second-order
model that was presented in Part~1.

To verify the proposed model,
we conducted simulations with typical detector data.
Compared with the first-order LWR model, the proposed second-order model
is more accurate.

The rest of the paper is organized as follows.
Section~\ref{sec-model} has two parts:
Subsection~\ref{subsec-system} describes the model;
Subsection~\ref{subsec-input-data} presents the algorithm for calibration
of the fundamental diagram.
Section~\ref{sec-results} discusses simulation results.
And, Section~\ref{sec-conclusion} concludes the paper.

\section{Model}\label{sec-model}
\subsection{System of equations}\label{subsec-system}
The Lighthill-Whitham-Richards (LWR) model approximates traffic flow as a
one-dimensional flow of incompressible
fluid~\citep{Lighthill1955,Richards1956}.
It assumes that
\begin{enumerate}
\item  there is functional dependency between traffic speed $v(t,x)$
and its density $\rho(t,x)$;
\item  the law of ``mass'' (the number of vehicles) conservation holds.
\end{enumerate}

We use the following notation: $\rho(t,x)$ is the number of vehicles
per unit length of roadway at time $t$ near the point with coordinate $x$;
$v(t,x)$ is the mean traffic speed at time $t$ near the point with coordinate
$x$.
The assumption of functional dependency between speed $v(t,x)$ and density
$\rho(t,x)$ can be expressed as:
\begin{equation} \label{v-of-rho} 
v(t,x) = V \left(\rho (t,x)\right),
\end{equation} 
where
\begin{equation} \label{v-prime-of-rho} 
V'(\rho)<0.
\end{equation}

The function $Q(\rho)=\rho V(\rho)$ specifying how vehicle flow
(number of vehicles crossing given cross-section in unit time)
depends on density is often called fundamental diagram
(in some publications, this term is reserved for $V(\rho)$).

The vehicle conservation law in the LWR model is expressed as a
differential form of continuity equation with the zero right-hand side:
\begin{equation} \label{model-lwr} 
\frac{\partial \rho}{\partial t} + \frac{\partial (\rho v)}{\partial x} = \frac{\partial \rho}{\partial t} + \frac{\partial \left(\rho V(\rho)\right)}{\partial x} =
\frac{\partial \rho}{\partial t} + \frac{\partial Q(\rho)}{\partial x} = 0.
\end{equation} 
Besides that, in the right-hand side of Eq.~\eqref{model-lwr}
we should account for possible variation in the number of vehicles due to
entrances and exits.
With this change, Eq.~\eqref{model-lwr} becomes:
\begin{equation} \label{model-lwr-rhs} 
\frac{\partial \rho}{\partial t} + \frac{\partial Q(\rho)}{\partial x} = f_{0}.
\end{equation} 

LWR model \citep{Lighthill1955,Richards1956,Whitham1974} and its
difference-differential and finite-difference analogs are popular
in applied modeling.
This is partially caused by distinct lack of available real-world data,
necessary for higher-order models 
(the accuracy improvements of higher-order models are offset by
inaccuracies in data).
A number of researchers focus on solving initial boundary value problems
for Eq.~\eqref{model-lwr-rhs} on a road network graph.
The main problems of this approach are in setting correct boundary conditions
for graph nodes.

The use of Eq.~\eqref{model-lwr-rhs} alone is not sufficient for correct
description of all phases of traffic flow \citep{Daganzo1995}.
We obtained the second-order macroscopic model in Part~1:
\begin{equation} \label{model} 
\left\{\begin{array}{l} 
\partial \rho / \partial t + \frac{\partial Q}{\partial x} = f_0, \\ 
\partial v / \partial t + \frac{\partial Q}{\partial \rho} \partial v / \partial x = \left( \frac{\partial Q}{\partial \rho} - v \right) \frac{f_{0}}{\rho}.
\end{array}
\right.  
\end{equation} 

We have also shown that according to the theorem from~\citep{Zhang2003}, 
our model guarantees anisotropy of traffic flow for solutions of
Eq.~\eqref{model}, since for all possible densities 
$0 \le \rho \le \rho_{*}$, the following holds for its eigenvalues: 
$\lambda_{1,2} = \frac{\partial Q(\rho)}{\partial \rho} < v(\rho)$.

In many second-order macroscopic models, the relaxation term 
$\frac{1}{\tau} \left(V(\rho)-v\right)$ is added to the right-hand side of
the momentum equation, which accounts for the drivers' desire to move
with the equilibrium speed $V(\rho)$.
It is usually assumed that $\tau \sim 1$ second is a typical driver
reaction time.
We did not include this reaction time in the model~\eqref{model}, because
all properties of traffic behavior are automatically accounted for in the
fundamental diagram $Q(\rho)$ that is constructed using traffic detector data.
To verify this approach, we conducted comparative simulations
with the relaxation term $\frac{1}{\tau} \left(V(\rho)-v\right)$
in the right-hand side of~\eqref{model}, and without it.
The results are discussed in the end of this paper.

\subsection{Input data}\label{subsec-input-data}
First, we determine the model parameters. 
To better describe the real-world situation, the model should account
for periodically updated detection data, road traffic regulations,
temporary road blocks, and so forth.

\begin{figure}[htb] 
 \includegraphics[width=\textwidth]{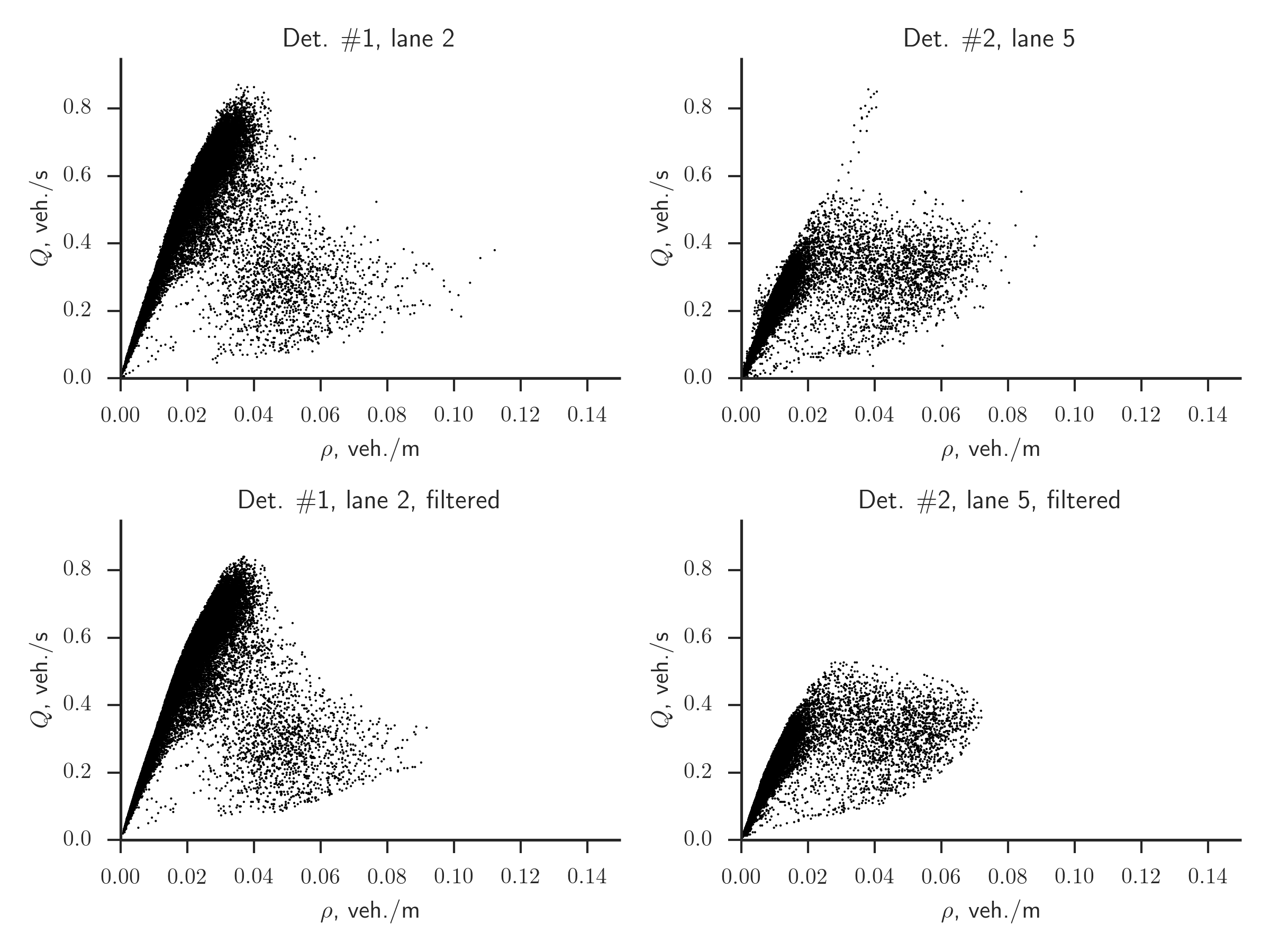}
 \caption{Observed data from two detectors installed in different lanes of the Moscow Ring Road --- measured flow $Q(\rho)$~veh./s for different densities $\rho$~veh./m. Measurements were made from January 1, 2012 to December 31, 2012, with 5-minute intervals. Top figures --- raw data; bottom figures --- data filtered through $\alpha$-hull peeling.}
\label{fig-input-data}
\end{figure}

We use data from stationary traffic detectors to construct fundamental diagrams
for all road segments.
Usually, this work is time-consuming due to unavailability of robust algorithms
for automatic fundamental diagram calibration based on traffic detector data.
For example, the widely used least-squares method for fitting empirical data
(e.g.~\citep{Wang2011}) suffers from poor fitting of data in congested regions.
This underlines the need for the accurate calibration algorithm.
Data from stationary freeway detectors allow us to parametrize
the fundamental diagram for a given road segment. 
The algorithm works as follows:
\begin{enumerate}
 \item Obtain data as pairs of values from a traffic detector
(sample data are shown in the top plots of Fig.~\ref{fig-input-data});
 \item Filter data using the $\alpha$-shape method~\citep{Edelsbrunner1983}.
First, we rescale to obtain similar scales for $Q$ and $\rho$ axes
(assuming that for a single lane, $Q$ values typically lie in the range
$[0;1]$~veh./s, and $\rho$ lies in the range $0;0.15$ veh./m,
we use rescaling coefficient 6.67).
Then we construct the $\alpha$-hull of resulting points,\footnote{The
$\alpha$-hull is a tight boundary (not necessarily convex) around a set of
points.}
and remove points belonging to the hull.
Then we construct $\alpha$-hull for the remaining points.
The process
(``onion-peeling'', originally proposed in~\citep{Eddy1982} for convex hulls)
is repeated until either less than 90\% of the original points remain,
or the relative difference in the hull area between the successive iterations
becomes less than 5\%.
Results are shown in the bottom plots in Fig.~\ref{fig-input-data};
 \item We set the maximal possible density $\rho_{\max}$ as the product of
number of lanes and constant 0.145~veh./m, which is based on a widely accepted
assumption that a vehicular movement becomes impossible with density
above 150~veh./km/lane: $\rho_{\max} =0.145$~veh./m/lane;
 \item Among the filtered data points, we determine the maximal flow value
$Q_1 \left(\rho_1 \right) = \mathop{\max}\limits_{\rho} \left[Q(\rho)\right]$
and the corresponding critical density $\rho_1$;
 \item We determine the intermediate point 
$Q_{0} \left(\rho_0 \right)=\mathop{\max}\limits_{Q} Q\left(\frac{\rho_1 }{2} \right)$ for some intermediate density $\rho_0 = \left(\frac{\rho_1}{2} \right)$;
 \item We find point farthest from the origin (after scaling) in 
$Q(\rho)$ plane: 
$Q_2 \left(\rho_2\right)= \max \sqrt{\left(\frac{Q}{Q_{\max}} \right)^2 + \left(\frac{\rho}{\rho_{\max}} \right)^2}$;
 \item After determining those key points, we build functional dependency $Q(\rho)$.
\end{enumerate}

According to the three-phase traffic theory~\citep{Kerner2009},
the following traffic phases are distinguished:
\begin{enumerate}
 \item Free flow: $Q\left(0 \le \rho < \rho_1 \right)$;
 \item Synchronized flow: $Q\left(\rho_1 \le \rho < \rho_2 \right)$;
 \item Wide moving jam: $Q\left(\rho_2 \le \rho \le \rho_{\max} \right)$.
\end{enumerate}

For each phase we define a separate function $Q(\rho)$,
stitching them together in the transition points:
\begin{enumerate}
 \item Free flow: $Q(\rho) = \alpha_2 \rho^2 + \alpha_1 \rho \mathrm{,\ } 0 \le \rho < \rho_1$;
 \item Synchronized flow: $Q(\rho) = \beta_2 \rho^2 + \beta_1 \rho + \beta_0 \mathrm{,\ } \rho_1 \le \rho < \rho_2$;
 \item Wide moving jam: $Q(\rho) = c_{*} (\rho_{\max} - \rho) \mathrm{,\ } \rho_2 \le \rho \le \rho_{\max}$.
\end{enumerate}

We determine the coefficients by forcing the function to be continuous through 
the following system of equations:
\begin{equation} \label{fund-cont} 
\left\{
 \begin{array}{l}
  \alpha_2 \rho_0^2 + \alpha_1 \rho_0 = Q(\rho_0); \\ 
  \alpha_2 \rho_1^2 + \alpha_1 \rho_1 = Q(\rho_1); \\ 
  \beta_2 \rho_1^2 + \beta_1 \rho_1 + \beta_0 = Q(\rho_1); \\ 
  \beta_2 \rho_2^2 + \beta_1 \rho_2 + \beta_0 = Q(\rho_2); \\ 
  2\beta_2 \rho_1 + \beta_1 = c_1 = \left. \frac{\partial Q(\rho)}{\partial \rho} \right|_{\rho =\rho_1}; \\ 
  c_{*} (\rho_{\max} - \rho_2) = Q(\rho_2).
 \end{array}
\right.  
\end{equation} 

\begin{figure}[htb] 
 \includegraphics[width=\textwidth]{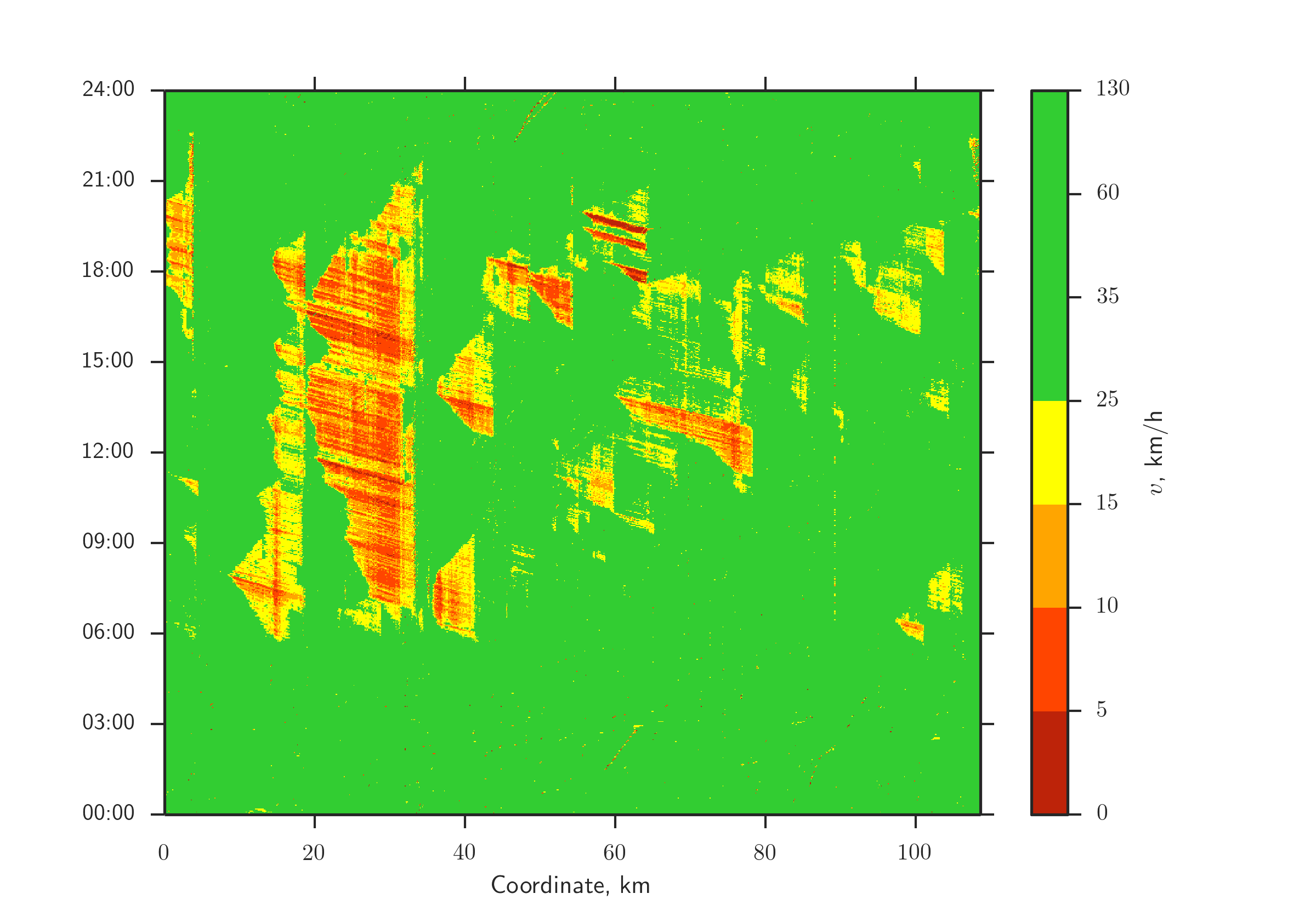}
 \caption{The space-time contour of traffic speed values for the Moscow Ring Road for a single weekday, December 5, 2012.}
\label{fig-mkad}
\end{figure}

The only missing value is that of the derivative of function to the right of
critical point:
$\left.\frac{\partial Q(\rho)}{\partial \rho} \right|_{\rho =\rho_1} = c_{1} =?$
in \eqref{fund-cont}.
Empirical observations suggest that traffic slows down significantly after
passing the critical point.
This leads to the assumption that $c_1$ is the speed of the deceleration wave,
and can be determined through the space-time speed contour for the roadway
segment in question (see Fig.~\ref{fig-mkad}).

\begin{figure}[htb]
 \includegraphics[width=\textwidth]{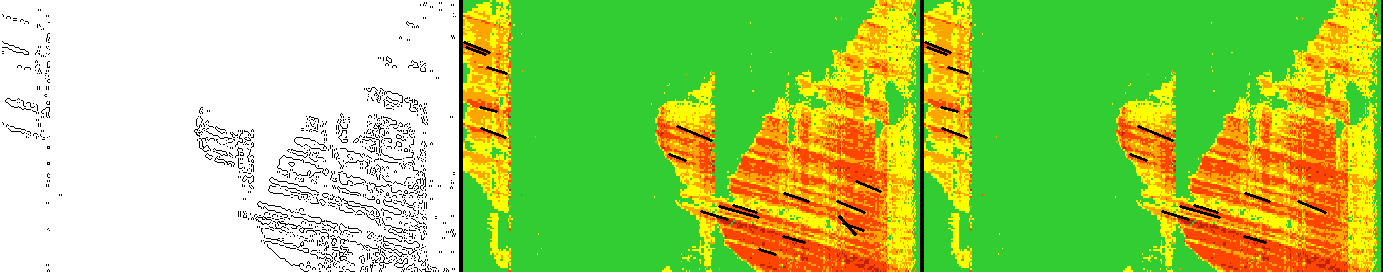}
 \caption{Automatic determination of the deceleration wave propagation speed from the space-time speed contour of the Moscow Ring Road.
Left: edge detection via Canny method.
Middle: segment detection using probabilistic Hough transform.
Right: segment filtration.}
\label{fig-algo}
\end{figure}

The red lines of deceleration waves propagating upstream are clearly visible
in Fig.~\ref{fig-mkad}.
The angle of these lines indicates the deceleration wave propagation speed
$c_1$.
We propose the algorithm for automatic determination of these lines from a given
speed contour.
The algorithm is based on computer vision methods implemented in OpenCV
library~\citep{Bradski2000}.
The algorithm has three steps: (see Fig.~\ref{fig-algo}):
\begin{enumerate}
 \item Edge detection using Canny algorithm~\citep{Canny1986} --- Fig.~\ref{fig-algo} (left);
 \item The detection of straight line segments in the contour using
probabilistic Hough transform \citep{Matas2000} --- Fig.~\ref{fig-algo}
(center);
 \item Filtering of the resulting segments: first, we count the number of data
points with $v < 10$~km/h in the segment and in its proximity; 
then we retain only segments having more than 65\% of their points with
$v < 10$~km/h;
such segments correspond to fronts of deceleration
waves --- Fig.~\ref{fig-algo} (right).
\end{enumerate}

\begin{figure}[htb] 
 \includegraphics[width=\textwidth]{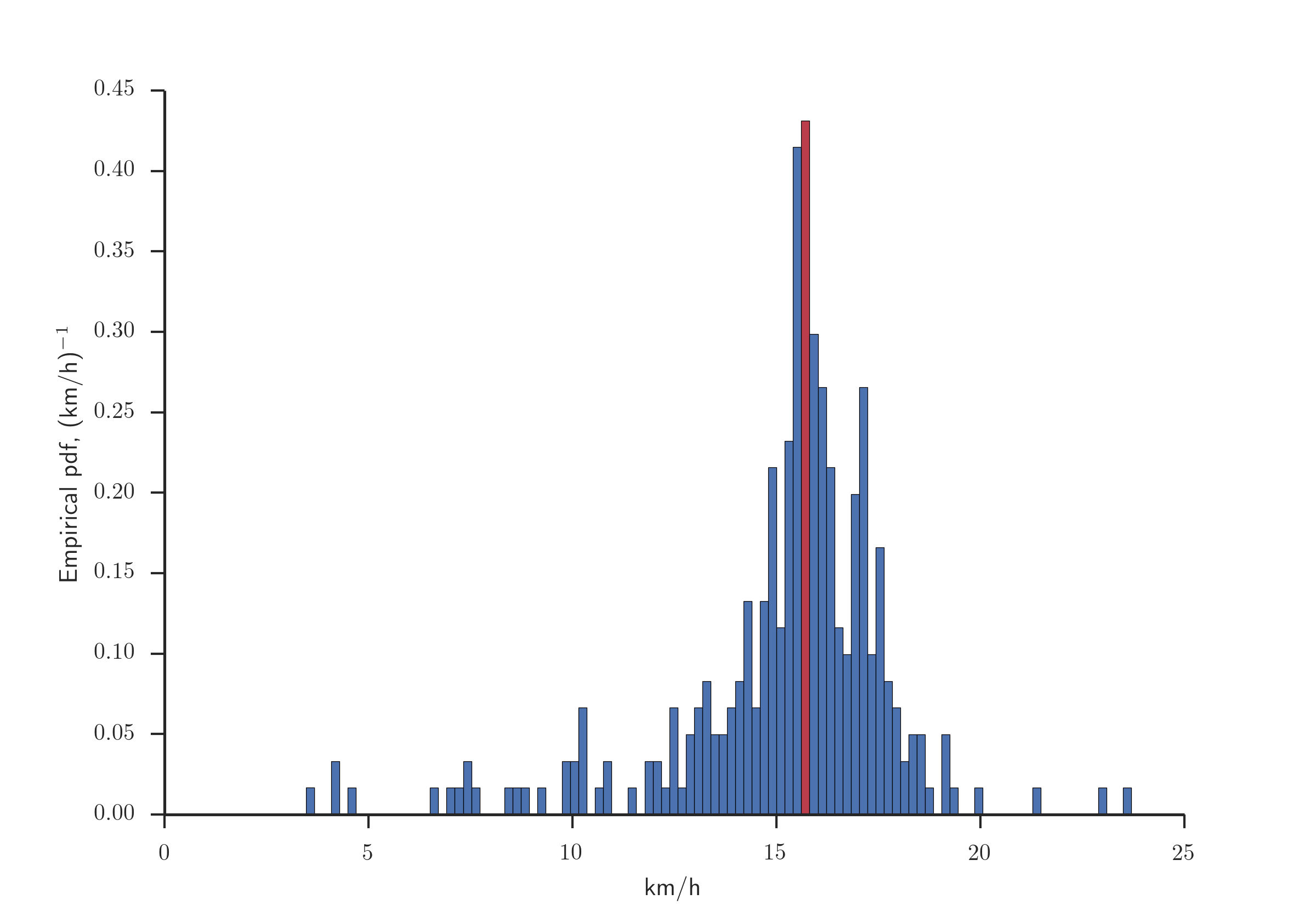}
 \caption{Histogram of absolute deceleration wave propagation speeds
for the Moscow Ring Road, constructed using data from year 2012.}
\label{fig-c-hist}
\end{figure}

As a result, we obtained the histogram of absolute deceleration wave
propagation speeds for the Moscow Ring Road (Fig.~\ref{fig-c-hist}).

We can see that the deceleration wave propagation speed on
the Moscow Ring Road is $\left. \frac{\partial Q(\rho)}{\partial \rho} \right|_{\rho =\rho_1} = c_1 =-15.8$~km/h.
We also determined that this speed is independent of the season or day of week,
and is solely determined by the road geometry.
It should be noticed that the deceleration wave propagation speed on
the Moscow Ring Road is very close in its value to the typical mean speed of
``the trailing edge of wide moving jam'' $v_g \approx -15$~km/h~\citep{Kerner2009}.
If GPS data are not available, $v_g$ could be used in system~\eqref{fund-cont}
instead of $c_1$.
Such approach was used in the first part of this paper,
since PeMS~\citep{pems} did not have GPS data for I-580.

For the traffic detector showcased in Fig.~\ref{fig-input-data},
we get the following results
(common values are $c_1 = 3.76$~m/s, $\rho_{\max} = 0.145$~veh./m):
\begin{itemize}
 \item Detector \#1: $Q_{1} = \max \left[Q(\rho)\right] = 0.84$~veh./s, $\rho_1 = 0.037$~veh./m; $Q_2 = \max \sqrt{\left(\frac{Q}{Q_{\max}} \right)^2 + \left(\frac{\rho}{\rho_{\max}} \right)^2} = 0.84$~veh./s, $\rho_2 = \rho_1 = 0.037$;
 \item Detector \#2: $Q_{1} = \max \left[Q(\rho)\right] = 0.59$~veh./s, $\rho_1 = 0.034$~veh./m; $Q_2 = \max \sqrt{\left(\frac{Q}{Q_{\max}} \right)^2 + \left(\frac{\rho}{\rho_{\max}} \right)^2} = 0.44$~veh./s, $\rho_2 = 0.076$~veh./m.
\end{itemize}

\begin{figure}[htb]  
 \includegraphics[width=\textwidth]{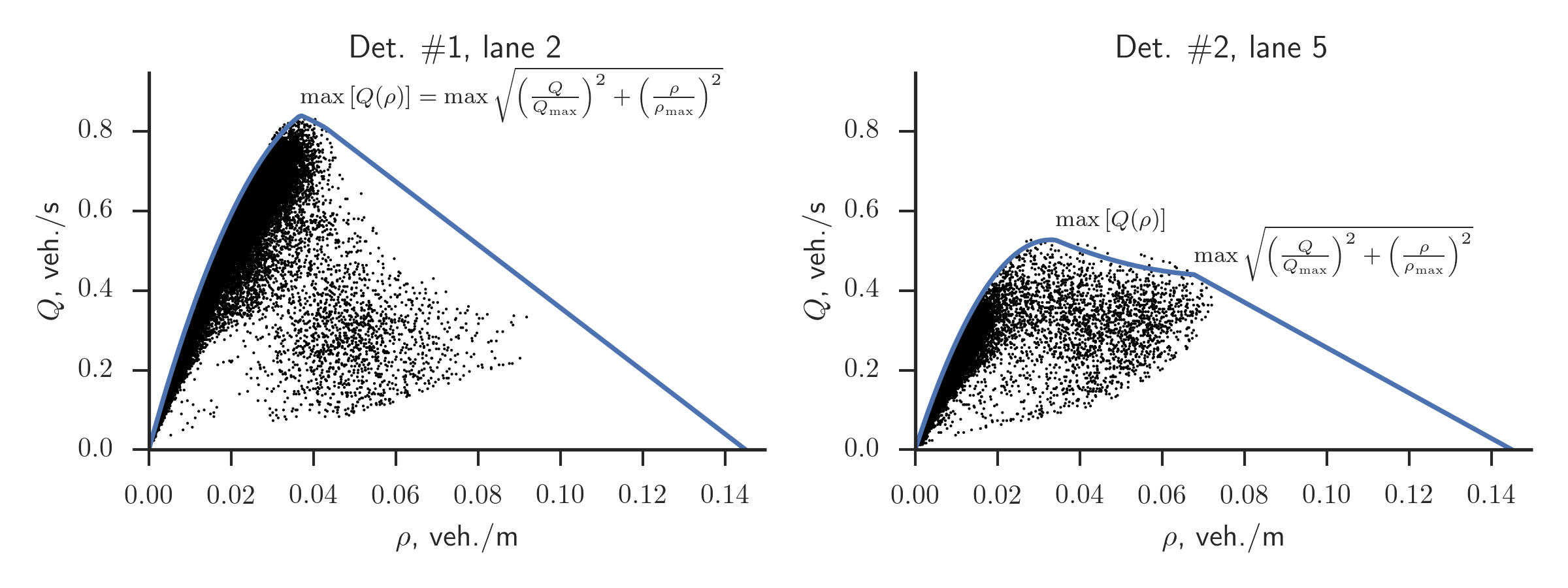}
 \caption{Fundamental diagrams for two points on the Moscow Ring Road.
Left: data from the second lane (detector \#~1);
right: data from the fifth lane (detector \#~2).}
\label{fig-fund-fit}
\end{figure}

We present the resulting fundamental diagrams in Fig.~\ref{fig-fund-fit}.
As we can see, the first diagram does not feature the synchronized flow stage,
present in the second case.

\begin{figure}[htb] 
 \includegraphics[width=\textwidth]{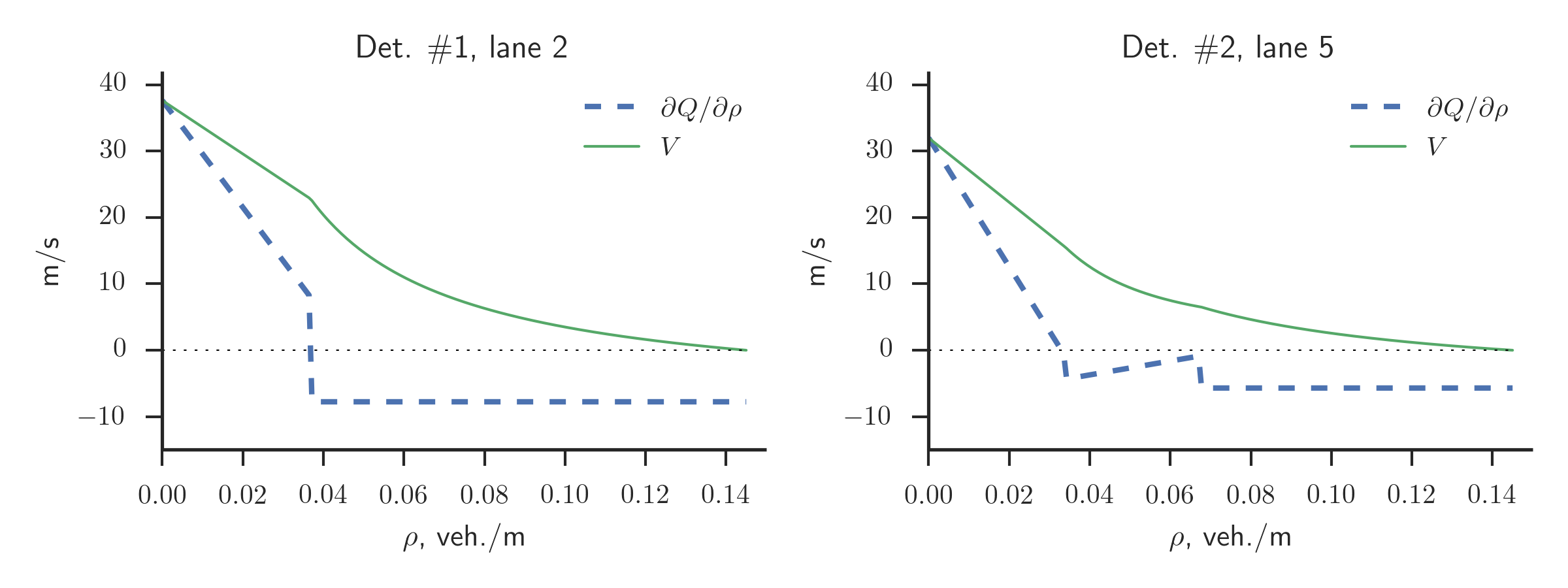}
 \caption{Eigenvalues $\lambda_{1,2} = \partial Q(\rho) / \partial \rho$ as
functions of density $\rho$ (dashed blue line),
compared to speed $v(\rho) = Q(\rho) / \rho$ (solid green line), 
for two fundamental diagrams, shown in Fig.~\ref{fig-fund-fit}. 
Left: data from the second lane (detector \#~1); 
right: data from the fifth lane (detector \#~2).}
\label{fig-fund-cv}
\end{figure}

If we plot eigenvalues of the system \eqref{model}: $\lambda_{1,2} = \frac{\partial Q(\rho)}{\partial \rho}$ as functions of density $\rho$, and compare them with equilibrium speed $v(\rho)=\frac{Q(\rho)}{\rho}$, as shown in Fig.~\ref{fig-fund-cv}, we can see that for all allowed density values ($0 \le \rho \le \rho_{\max}$), it holds true that $\lambda_{1,2} = \frac{\partial Q(\rho)}{\partial \rho} < v(\rho)$. This means, according to theorem proven in \citep{Zhang2003}, that traffic flow is anisotropic on studied regions of the Moscow Ring Road.

\begin{figure}[htb]  
 \includegraphics[width=\textwidth]{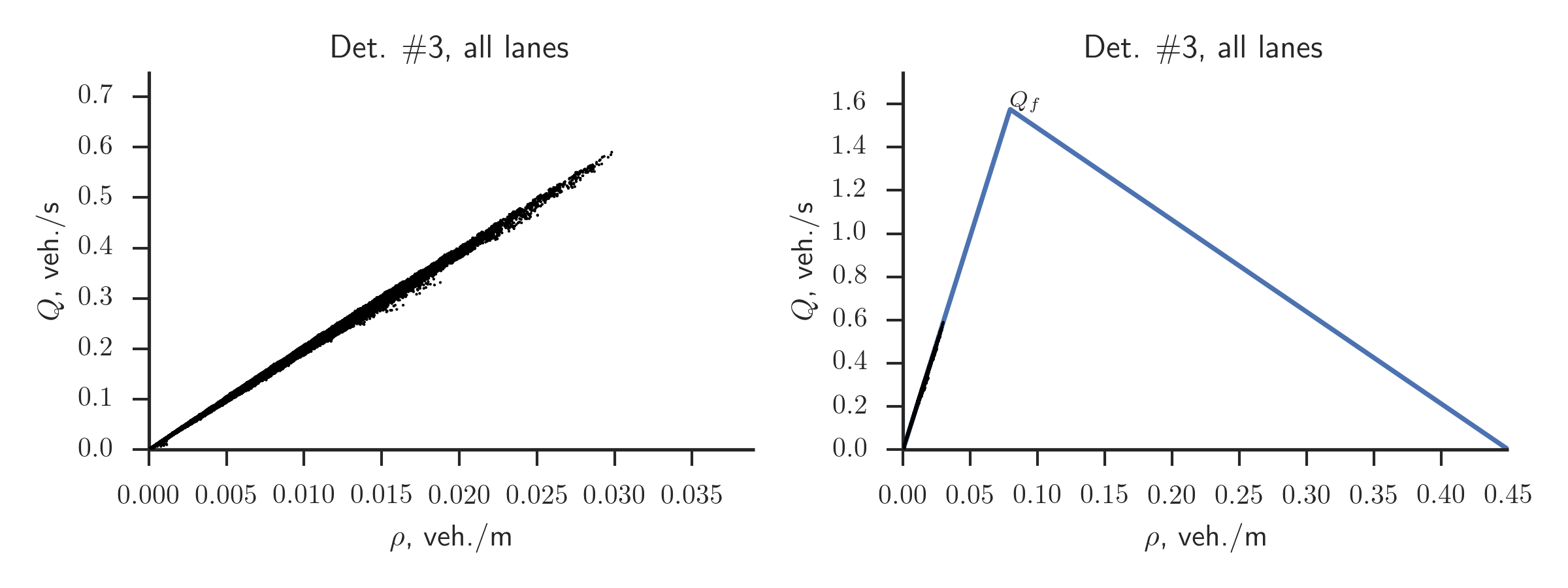}
 \caption{Empirical data for a segment of Mamadyshskiy Trakt, as measured from February to September 2014, with 5-minute measuring interval.
Left: filtered data; right: fundamental diagram built using these data.}
\label{fig-fund-kazan}
\end{figure}

Now we discuss what to do if measurement data are not sufficient.
As an example, we study data from traffic detector installed on
Mamadyshskiy Trakt --- a 3-lane wide intercity freeway near Kazan, Russia.
(Fig.~\ref{fig-fund-kazan}).

As we can see from Fig.~\ref{fig-fund-kazan} (left), Mamadyshskiy Trakt is well below its capacity, and the data do not cover the whole range of possible
density values $0\le \rho \le \rho_{\max} = 0.45$.
Therefore, we assume that the maximal flow for a single lane is
0.525~veh./s = 1890~veh./hour
(usually, the maximal flow value for a single lane is assumed to be
between 1800 and 2000~veh./hour), and $Q_f = \mathop{\max}\limits_{\rho} \left[Q(\rho)\right] = 0.525 \times 3 = 1.575$~veh./s.
Using this value, we calibrate the fundamental diagram as follows:
\begin{enumerate}
 \item We determine the maximal flow value present in data
$Q_1 (\rho_1) = \mathop{\max}\limits_{\rho} \left[Q(\rho)\right]$
and the corresponding density $\rho_1$,
which in this case is not critical;
 \item We determine the intermediate point
$Q_0 (\rho_0) = \mathop{\max}\limits_{Q} Q\left(\frac{\rho_1}{2} \right)$
for the intermediate density $\rho_0 = \frac{\rho_1}{2}$;
 \item We find coefficients for the free-flow part of the fundamental diagram
$\left\{\alpha_1, \alpha_2 \right\}$ by passing a second-degree polynomial
through the following points:
\begin{equation} \label{a1-a2} 
\left\{
 \begin{array}{l}
  \alpha_2 \rho_0^2 + \alpha_1 \rho_0 = Q(\rho_0), \\ 
  \alpha_2 \rho_1^2 + \alpha_1 \rho_1 = Q(\rho_1).
 \end{array}
\right.
\end{equation} 
 \item We find the critical density by solving the quadratic equation:
$\alpha_2 \rho_f^2 + \alpha_1 \rho_f = Q_f$.
 \item We find braking speed $c_f$ from the ratio
$c_f = \frac{Q_f}{\left(\rho_{\max} - \rho_f \right)}$.
 \item We obtain the final form of the fundamental diagram $Q(\rho)$
for two phases of traffic: free
$Q(\rho) = \alpha_2 \rho^2 + \alpha_1 \rho \textrm{, } 0 \le \rho < \rho_f$ and
wide moving jam
$Q(\rho) = c_f \left(\rho_f - \rho \right) \textrm{, } \rho_f \le \rho \le \rho_{\max}$
(see Fig.~\ref{fig-fund-kazan}).
In this case, we ignore the synchronized flow phase.
\end{enumerate}

\section{Results}\label{sec-results}
\begin{figure}[htb]  
 \includegraphics[width=\textwidth]{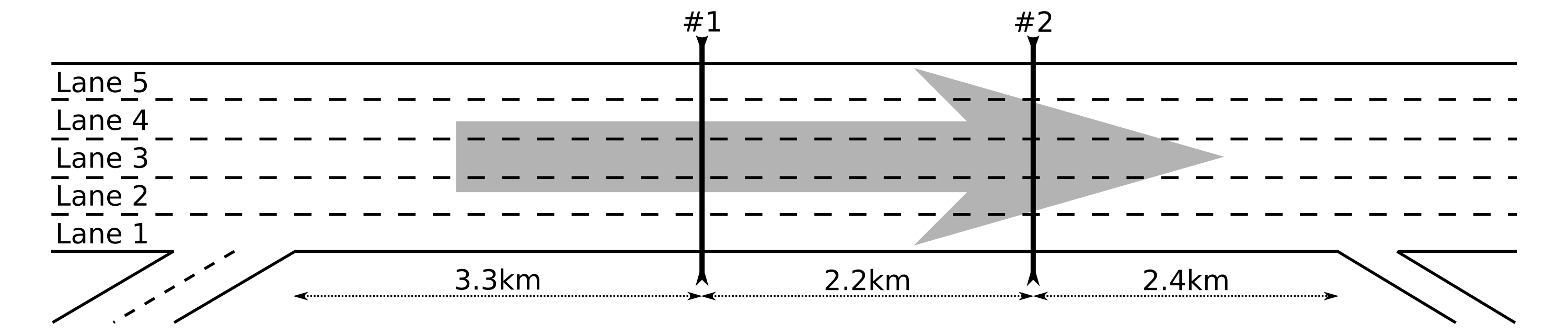}
 \caption{The segment of the Moscow Ring Road containing traffic detectors \#~1 and \#~2.
We used data from detector \#~1 (flow and speed) as boundary conditions,
and verified the numerical results against data from the
downstream detector \#~2.}
\label{fig-scheme}
\end{figure}

\begin{figure}[htb]  
 \includegraphics[width=\textwidth]{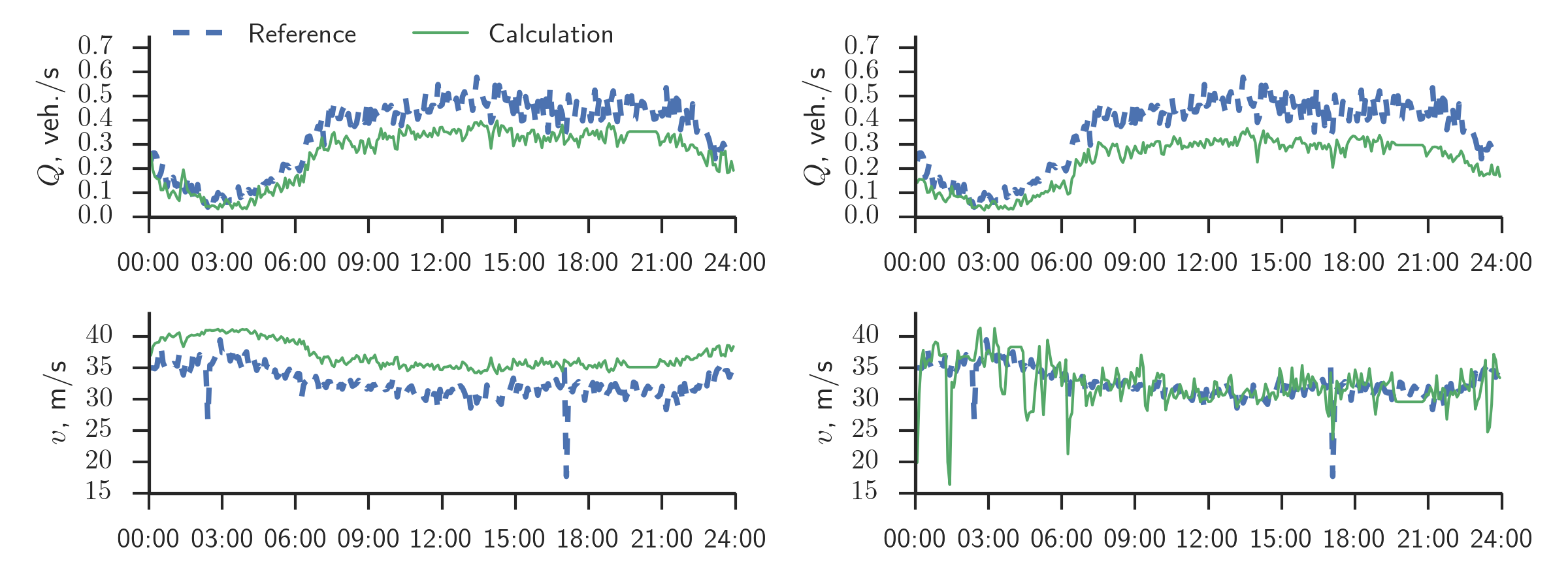}
 \caption{Comparison of calculated flows $Q$ and speeds $v$ (solid green line), with respective data from detector \#~2 (dashed blue line) for the first lane.
Left: the results obtained with the first-order LWR model~\eqref{model-lwr-rhs};
middle: the results obtained with the proposed second-order model~\eqref{model}
with the relaxation term;
right: the results obtained with model~\eqref{model} without the relaxation term.}
\label{fig-res-lane1}
\end{figure}

\begin{figure}[htb]  
 \includegraphics[width=\textwidth]{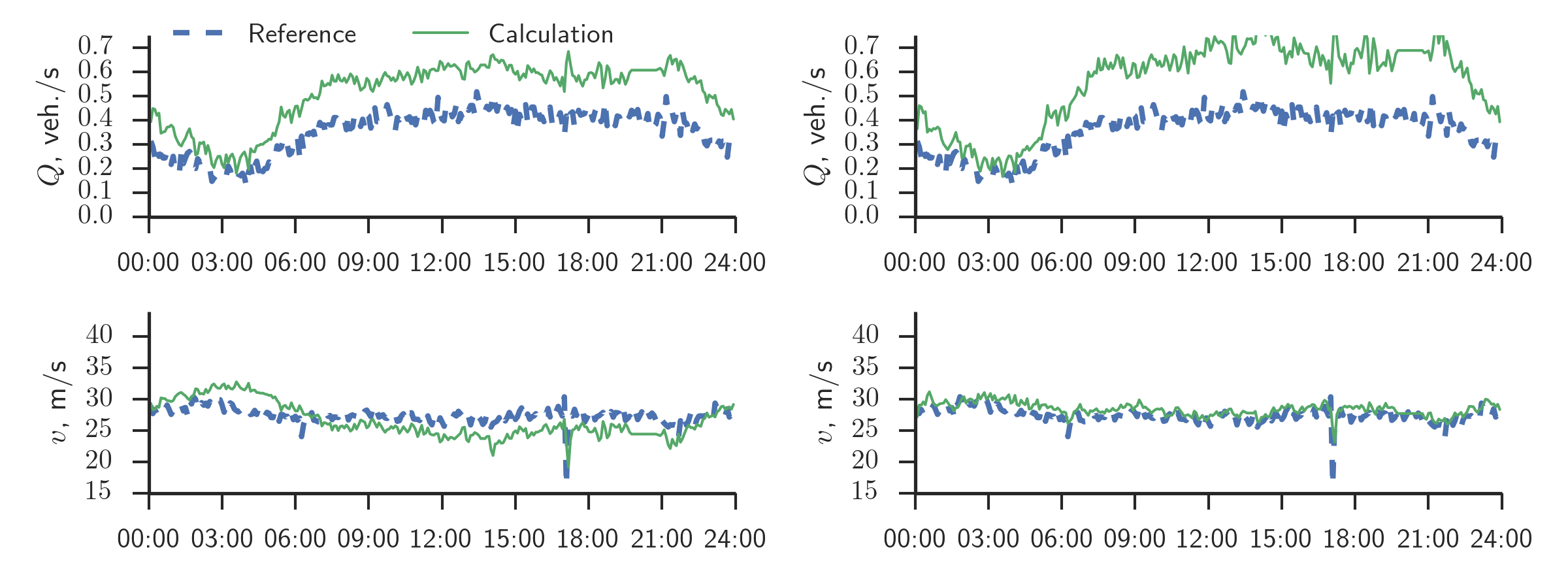}
 \caption{Comparison of calculated flows $Q$ and speeds $v$ (solid green line), with respective data from detector \#~2 (dashed blue line) for the second lane.
Left: the results obtained with the first-order LWR model~\eqref{model-lwr-rhs};
middle: the results obtained with the proposed second-order model~\eqref{model}
with the relaxation term;
right: the results obtained with model~\eqref{model} without the relaxation term.}
\label{fig-res-lane2}
\end{figure}

\begin{figure}[htb]  
 \includegraphics[width=\textwidth]{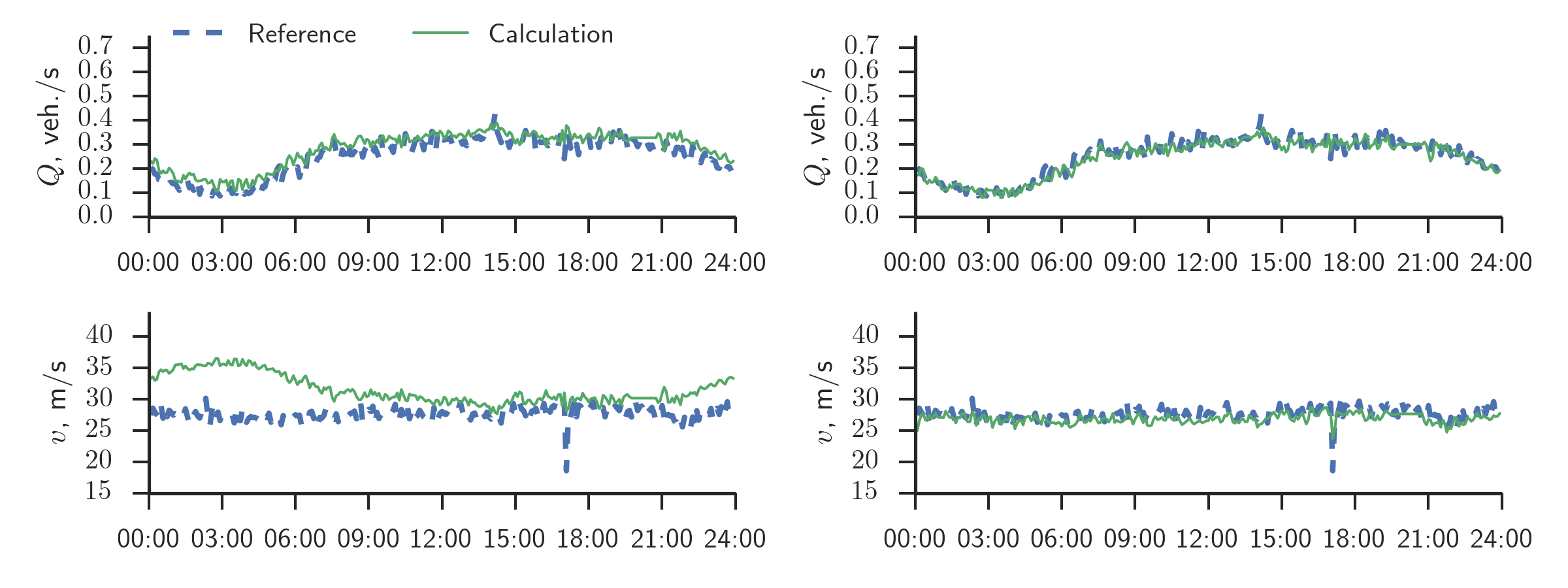}
 \caption{Comparison of calculated flows $Q$ and speeds $v$ (solid green line), with respective data from detector \#~2 (dashed blue line) for the third lane.
Left: the results obtained with the first-order LWR model~\eqref{model-lwr-rhs};
middle: the results obtained with the proposed second-order model~\eqref{model}
with the relaxation term;
right: the results obtained with model~\eqref{model} without the relaxation term.}
\label{fig-res-lane3}
\end{figure}

\begin{figure}[htb]  
 \includegraphics[width=\textwidth]{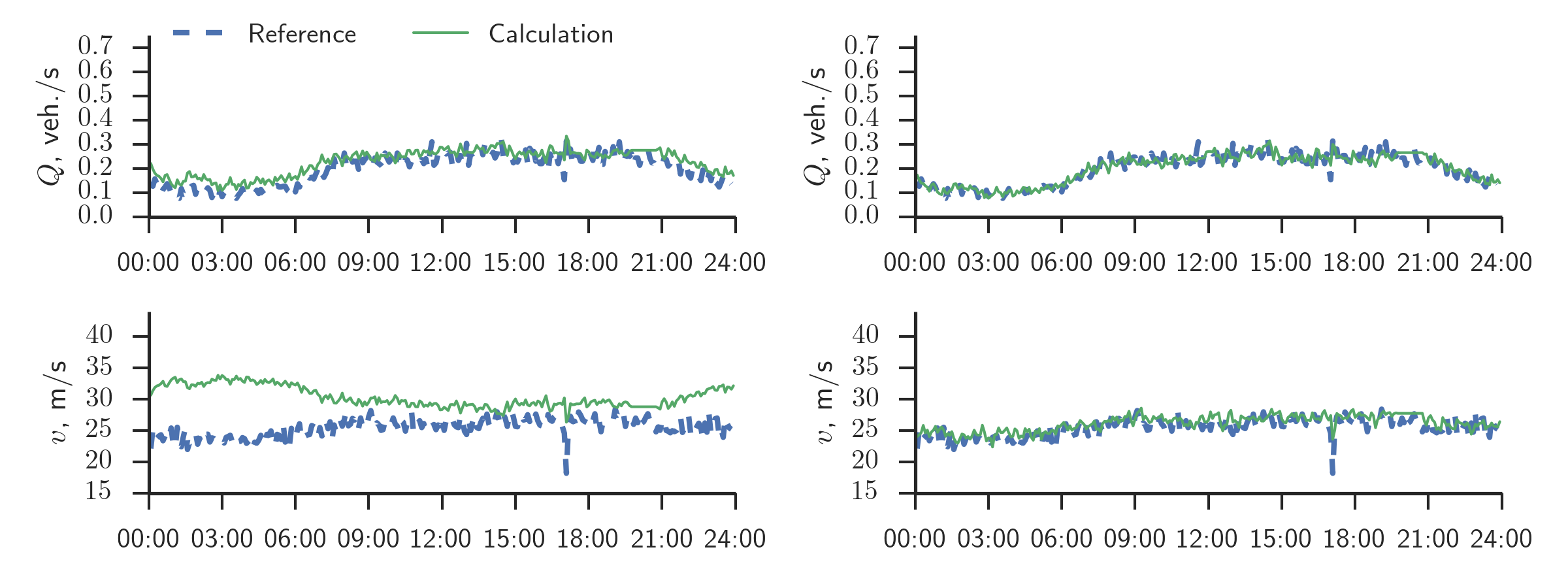}
 \caption{Comparison of calculated flows $Q$ and speeds $v$ (solid green line), with respective data from detector \#~2 (dashed blue line) for the fourth lane. 
Left: the results obtained with the first-order LWR model~\eqref{model-lwr-rhs};
middle: the results obtained with the proposed second-order model~\eqref{model}
with the relaxation term;
right: the results obtained with model~\eqref{model} without the relaxation term.}
\label{fig-res-lane4}
\end{figure}

\begin{figure}[htb]  
 \includegraphics[width=\textwidth]{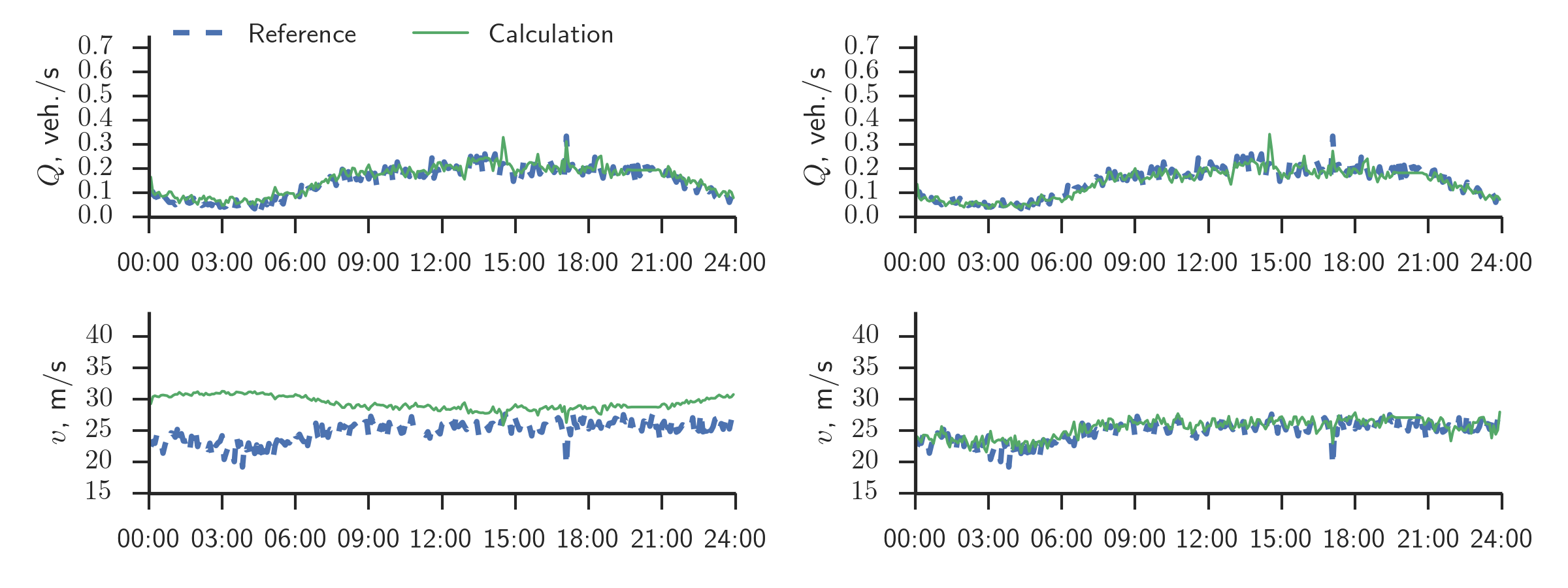}
 \caption{Comparison of calculated flows $Q$ and speeds $v$ (solid green line), with respective data from detector \#~2 (dashed blue line) for the fifth lane.
Left: the results obtained with the first-order LWR model~\eqref{model-lwr-rhs};
middle: the results obtained with the proposed second-order model~\eqref{model}
with the relaxation term;
right: the results obtained with model~\eqref{model} without the relaxation term.}
\label{fig-res-lane5}
\end{figure}

\begin{figure}[htb] 
 \includegraphics[width=\textwidth]{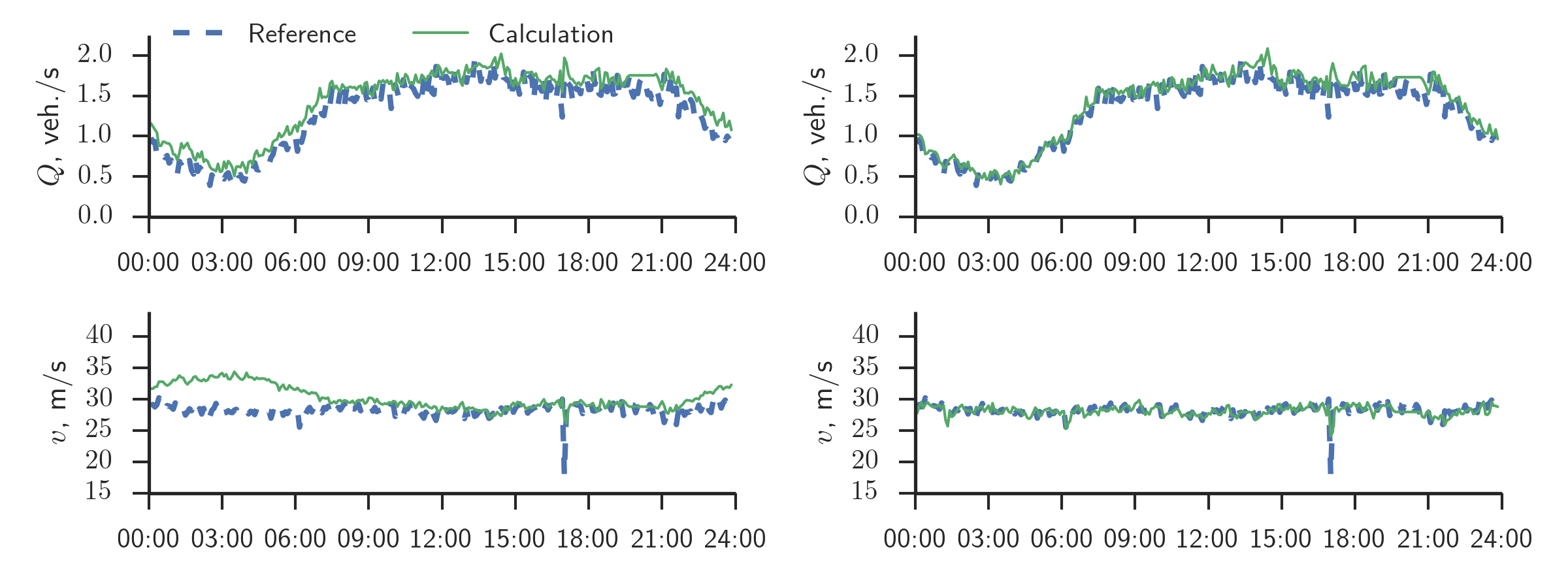}
 \caption{Comparison of calculated flows $Q$ and speeds $v$ (solid green line), with respective data from detector \#~2 (dashed blue line), aggregated over all lanes. 
Left: the results obtained with the first-order LWR model~\eqref{model-lwr-rhs};
middle: the results obtained with the proposed second-order model~\eqref{model}
with the relaxation term;
right: the results obtained with model~\eqref{model} without the relaxation term.}
\label{fig-res-lane-all} 
\end{figure}

To verify the proposed approach of using automatic fundamental diagram
estimation algorithm, we conducted numerical experiment for one of the segments
of the Moscow Ring Road (see Fig.~\ref{fig-scheme}).
For this, we used data from two traffic detectors, denoted
\#~1 and \#~2, with distance 2.2~km between them, and no entrances or exits.
We used data from detector \#~1 (flow and speed) as boundary conditions
and verified the numerical results against data from the downstream detector
\#~2.
In simulation, we use transparent boundary conditions at the location
of detector \#~2 in:
$\partial Q / \partial x = 0$,  $\partial v / \partial x = 0$.
We simulated 24 hours of a weekday.
Results are shown in Figs.~\ref{fig-res-lane1}--\ref{fig-res-lane5}
for each lane separately, and the aggregated data over all lanes are
shown in Fig.~\ref{fig-res-lane-all}.
Left subfigures show the results obtained with the first-order LWR
model~\eqref{model-lwr-rhs};
the middle subfigures --- with the proposed anisotropic model~\eqref{model}
with relaxation term $\frac{1}{\tau} \left(V(\rho)-v\right)$ in the
right-hand side; 
the right subfigures --- with model~\eqref{model} without the relaxation term.

The simulation results demonstrate correct behavior of the proposed second-order anisotropic model.
The deviations of calculated flows with the observed ones in the first and 
the second lanes
(top plots in Figures~\ref{fig-res-lane1}--\ref{fig-res-lane2} are due to an
off-ramp after detector \#~2, which causes active lane changes between the first
two lanes.
We see that the calculated flows in the first lane are lower than the
observed ones (Fig.~\ref{fig-res-lane1}), and the opposite is true for the
second lane (Fig.~\ref{fig-res-lane2}).
There results confirm that drivers perform lane-change maneuvers from the first 
to the second lanes.
We did not account for lane changes when modeling individual lanes,
yet the simulation of all lanes together (Fig.~\ref{fig-res-lane1})
show correct results.
We also see the advantages of the proposed second-order anisotropic 
model~\eqref{model} over LWR model~\eqref{model-lwr-rhs}.
Although flows generated by different models are almost identical, 
speeds obtained with model~\eqref{model} agree with measurements much better
than those obtained with the LWR model~\eqref{model-lwr-rhs}.
Also, we should note that adding the relaxation term 
$\frac{1}{\tau} \left(V(\rho)-v\right)$ to the right-hand side of 
the momentum equation leads to the loss of accuracy for traffic speed.
This can be seen in all Figs.~\ref{fig-res-lane1}--\ref{fig-res-lane-all}.

\begin{figure}[htb] 
 \includegraphics[width=\textwidth]{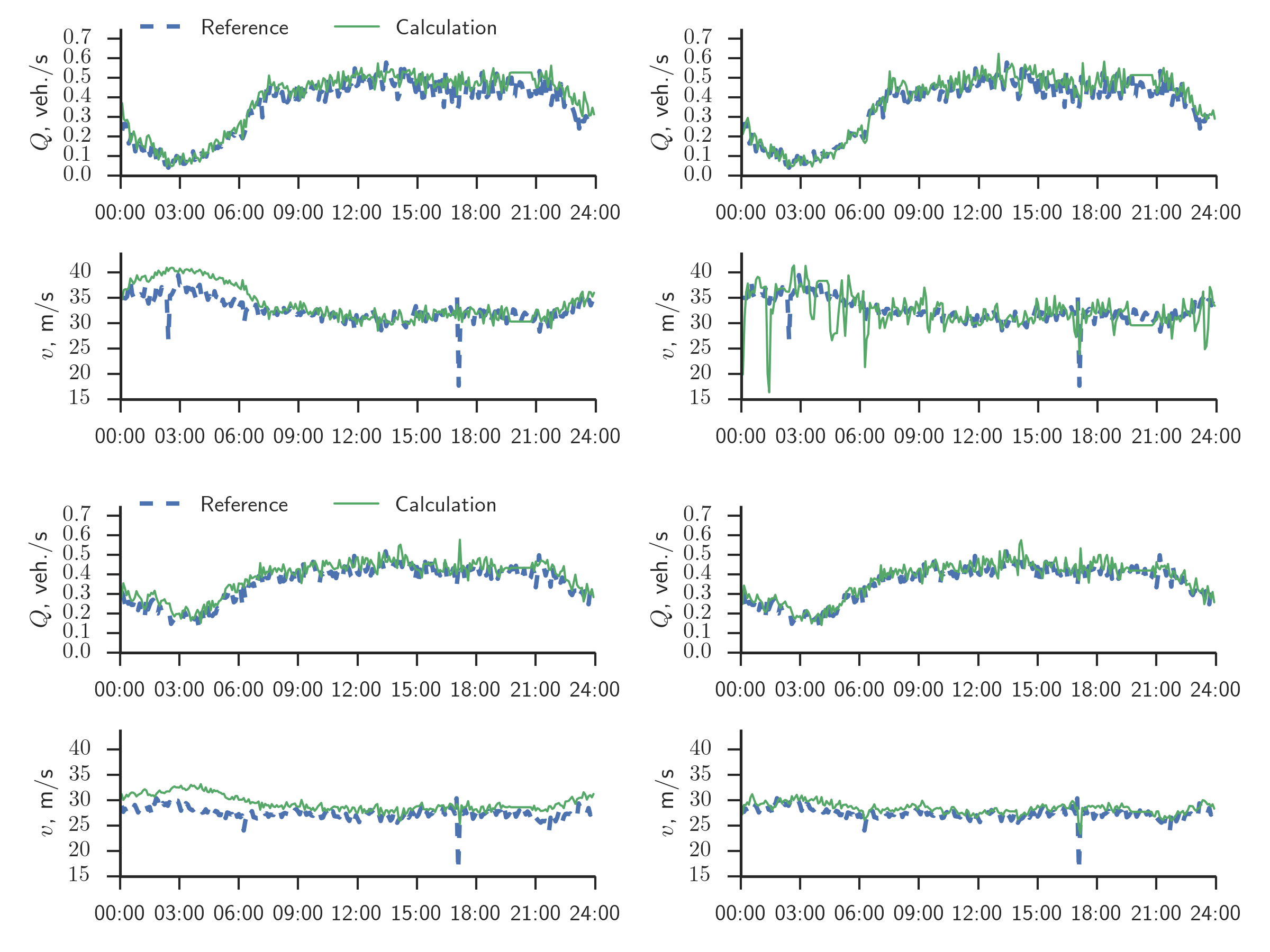}
 \caption{Comparison of calculated flows $Q$ and speeds $v$ (solid green line), with respective data from detector \#~2 (dashed blue line), for the first and the second lanes, accounting for lane changes. 
Left: the results obtained with the first-order LWR model~\eqref{model-lwr-rhs};
middle: the results obtained with the proposed second-order model~\eqref{model}
with the relaxation term;
right: the results obtained with model~\eqref{model} without the relaxation term.}
\label{fig-res-lane-weave} 
\end{figure}

If we take into account the lane changes between the first two lanes through
the right-hand side of the conservation equation in~\eqref{model-lwr-rhs}
and~\eqref{model}, then the computed flows for each lane will match
the observed ones significantly better (see Fig.~\ref{fig-res-lane-weave}).

\section{Conclusion}\label{sec-conclusion}
In this paper, we studied the problem of automatic calibration of macroscopic 
second-order traffic models.
Traffic flows and densities were obtained empirically for each segment of
the roadway through algorithm for processing data from stationary detectors
and GPS traces.

We verified the model proposed in the first part of the paper and its
calibration method with numerical experiments using traffic measurements
from the Moscow Ring Road.
The numerical results show the correct behavior of the proposed model,
and its advantage in precision over the
first-order LWR model~\citep{Lighthill1955,Richards1956,Whitham1974}
with respect to traffic speed.

We also studied the use of the relaxation term
$\frac{1}{\tau} \left(V(\rho)-v\right)$
in the right-hand side of the momentum equation.
It turns out that all necessary features of drivers' behavior can be taken into
account in the estimation of the fundamental diagram.
This means that adding the relaxation term
$\frac{1}{\tau} \left(V(\rho)-v\right)$ to the right-hand side of
the momentum equation is leads to the loss of precision in calculation of speed
and should be avoided.

\section{References}
\bibliographystyle{elsarticle-harv}
\bibliography{TransportationResearch}

\section{Appendix}\label{sec-appendix}

\subsection{Canny edge detector}\label{subsec-canny}

Canny edge detection algorithm \citep{Canny1986} is a robust method for detecting edges in grayscale images. The algorithm consists of the following steps:

\begin{itemize}
 \item \textbf{Denoising}: Since edge detection results can be significantly altered by noise, a $5 \times 5$ Gaussian filter is applied to smooth and denoise image.
 \item \textbf{Intensity Gradient}: The horizontal and vertical Sobel kernels (with aperture 5) are applied to the image to get first derivatives in horizontal ($G_x$) and vertical ($G_y$) directions. Then, edge angle $\theta = \tan ^{-1}\left(G_y / G_x\right)$ and gradient $G = \sqrt{G_x^2 + G_y^2}$ are determined for each pixel. The angle is further rounded to one of four possible values, corresponding to horizontal, vertical, or one of two possible diagonal directions.
 \item \textbf{Non-maximum suppression}: The edge strength for each pixel is compared to the strengths of its neighbours in the direction perpendicular to edge (for example, if the pixel is supposed to lie on a vertical edge, its value is compared to values of its left and right neighbours). If the pixel value is not the largest, it is set to zero (suppressed). This step is necessary for obtaining image with thin edges, since otherwise edges obtained from step (2) are still blurred.
 \item \textbf{Hysteresis thresholding}: In this step, we remove noise edges. For each pixel, if its gradient $G$ is above chosen upper threshold, then is is cosidered to be a ``sure-edge'', and if it below lower threshold, it is discarded. For values inbetween, we check whether they are connected to any of ``sure-edge'' pixels, and if so, they are considered a valid edge, and discarded otherwise.
\end{itemize}

In our case, we used binarized velocity map (with threshold set to 15~km/h) as input matrix.

\subsection{Progressive probabilistic Hough transform}\label{subsec-hough}

In classical Hough transform for straight line detection \citep{Hough1959,Duda1972}, the input image is mapped into Hough space $(r, \theta)$ of all possible lines on the plane (when written as $r = x \cos \theta + y \sin \theta$).
This way, each pixel $(x,y)$ in input binary image is transformed into sinusoid curve in $(r, \theta)$ space, which corresponds to all possible lines passing through this pixel. 
If the sinusoids of two different pixels intersect in $(r, \theta)$ plane, this means that these two pixels lie on the same line. 
The Hough Line transform calculates the number of intersections in each point of $(r, \theta)$ plane (so-called voting step), and considers a line detected if this number exceeds chosen threshold.
This is typically implemented using accumulator array for $(r, \theta)$ space with chosen steps in $r$ and $\theta$ dimensions, and iterating over all edge pixels of the image.

In probabilistic Hough transform (see, e.g., \citep{Kiryati1991}), only a number of edge points are used in voting step in order to improve algorithm speed.
However, this requires \textit{a priori} choice of sample size.
Progressive probabilistic Hough transform \citep{Matas2000} works by checking whether any bin exceeds dynamically calculated significance threshold $l$ after sampling each edge point. With some optimizations, the algorithm works as follows:

\begin{itemize}
 \item Until input image is empty:
 \begin{itemize}
 \item \textbf{Sample}: Remove random edge pixel from image and update accumulator array in $(r, \theta)$ space.
 \item \textbf{Check peak}: If the highest peak in accumulator array exceeds chosen threshold $T$:
 \begin{itemize}
  \item \textbf{Find segment}: Check the line corresponding to peak and find the longest segment that has no gaps exceeding given threshold $G$. The pixels belonging to the segment are removed from image, and their votes, if any, are removed from accumulator.
  \item \textbf{Save output}: If length of the segment exceeds chosen threshold $L$, it is added to output list.
 \end{itemize}
\end{itemize}
\end{itemize}

\end{document}